\documentclass[a4paper,12pt]{article}
\usepackage[dvips]{graphicx}
\begin{document}
\begin{center}
\Large\bf
Local thermalization in the d + Au system\\[2.1cm]
\large\rm
Georg Wolschin$^{1,2}$, Minoru
Biyajima$^{2}$,\\ Takuya Mizoguchi$^{3}$, and Naomichi Suzuki$^{4}$\\[.8cm]
\normalsize\sc\rm
$^{1}$ Theoretical Physics, Heidelberg University,
D-69120 Heidelberg, Germany\\
$^{2}$ Department of Physics, Shinshu University, Matsumoto 390-8621,
Japan\\
$^{3}$ Toba National College of Maritime Technology, Toba 517-8501, Japan\\
$^{4}$ Department of Comprehensive Management,\\ Matsumoto University,
Matsumoto 390-1295,
Japan\\[2.6cm]
\end{center}
\bf
Abstract\\
\rm
The extent of a locally equilibrated parton plasma in d + Au collisions
at $\sqrt{s_{NN}}$ = 200 GeV is investigated as a 
function of collision centrality in a nonequilibrium-statistical 
framework. Based on a three-sources model, analytical solutions
of a relativistic diffusion equation are in precise agreement with recent 
data for charged-particle pseudorapidity distributions. 
The moving midrapidity source indicates the size of the local
thermal equilibrium region after hadronization. In central d + Au 
collisions it contains about 19\% of the produced particles, and
its relative importance rises with decreasing centrality.
\\[.4cm]
\newpage
The production and identification of a transient quark-gluon plasma
in local thermal equilibrium is of basic importance in relativistic
heavy-ion physics. In this Letter we propose nonequilibrium-statistical 
methods to investigate analytically the gradual thermalization occuring in the
course of particle production at the highest available energies. The
approach is tailored to identify the fraction of produced particles in local 
thermal equilibrium from their distribution functions in pseudorapidity.
It yields indirect evidence for the extent of a locally equilibrated parton plasma. 

Recently pseudorapidity
distributions of primary charged particles have become available
\cite{bbb05} as functions of centrality in d + Au collisions at
a nucleon-nucleon center-of-mass energy of 200 GeV. They are investigated
within a nonequilibrium-statistical framework that is based on
analytical solutions of a Relativistic Diffusion Model (RDM).

Our investigation is based on a linear  
Fokker-Planck equation (FPE) 
for three components $R_{k}(y,t)$ of the distribution function
in rapidity space
\cite{wol99,biy02,wol03} 

\begin{equation}
\frac{\partial}{\partial t}R_{k}(y,t)=
\frac{1}{\tau_{y}}\frac{\partial}
{\partial y}\Bigl[(y-y_{eq})\cdot R_{k}(y,t)\Bigr]
+\frac{\partial^2}{\partial^{2} y}\Bigl[D_{y}^{k}
\cdot R_{k}(y,t)\Bigr]
\label{fpe}
\end{equation}\\
with the rapidity $y=0.5\cdot ln((E+p)/(E-p))$.
The diagonal components $D_{y}^{k}$ of the diffusion tensor  
contain the microscopic
physics in the respective Au-like (k=1), d-like (k=2)
and central (k=3) regions. They 
account for the broadening of the distribution 
functions through interactions and particle creations. 
In the present investigation the off-diagonal terms of the
diffusion tensor are assumed to be zero.
The rapidity relaxation time $\tau_{y}$ determines
the speed of the statistical equilibration in y-space.

As time goes to infinity, the mean values of the
solutions of Eqs. (\ref{fpe}) approach the equilibrium value $y_{eq}$. 
We determine it 
from energy- and momentum conservation \cite{bha53,nag84}
in the system of Au- and d-participants and hence, it 
depends on impact parameter. This dependence is decisive 
for a detailed description of the measured charged-particle
distributions in asymmetric systems:

\begin{equation}
y_{eq}(b)=1/2\cdot ln\frac{<m_{1}^{T}(b)>exp(y_{max})+<m_{2}^{T}(b)>
exp(-y_{max})}
{<m_{2}^{T}(b)>exp(y_{max})+<m_{1}^{T}(b)>exp(-y_{max})}
\label{yeq}
\end{equation}\\
with the beam rapidities y$_{b} = \pm y_{max}$, the transverse
masses $<m_{1,2}^{T}(b)>=\\
\sqrt(m_{1,2}^2(b)+<p_{T}>^2)$, and masses
m$_{1,2}(b)$ of the Au- and d-like participants 
that depend on the impact parameter $b$. The average 
numbers of participants $N_{1,2}(b)$
in the incident gold and deuteron nuclei are calculated from the
geometrical overlap. The results are consistent with the Glauber
calculations reported in \cite{bbb05} which we use in the further
analysis. The corresponding equilibrium values of the rapidity
vary from y$_{eq}=$ - 0.169 for peripheral (80-100$\%$) to 
y$_{eq}=$ - 0.944 for central (0-20$\%$) collisions.
They are negative due to the net longitudinal momentum of the
participants in the laboratory frame, and their absolute
magnitudes decrease with impact parameter since the number of
participants decreases for more peripheral collisions.

The RDM describes the drift towards $y_{eq}$ in a statistical sense.
Whether the mean values of the distribution functions $R_{1}$ and 
$R_{2}$ actually attain $y_{eq}$ depends on the interaction time 
$\tau_{int}$ (the time the system interacts strongly, or the integration time
of (\ref{fpe})). It can be determined from dynamical models or
from parametrizations of two-particle correlation measurements. For
central Au + Au at 200 A GeV, this yields about
$\tau_{int}\simeq 10 fm/c$ \cite{lis05}, which is too short for $R_{1}$ and 
$R_{2}$ to reach equilibrium. Note, however, that this does
not apply to $R_{eq}$ which is born near local equilibrium at short 
times (in the present calculation, at t = 0 due to the
$\delta-$function intitial conditions),
and then spreads in time through diffusive interactions with other
particles at nearly the same rapidity. 

Our analytical diffusion model is consistent with, and 
complementary to parton cascade models
where stopping involves large sudden jumps in rapidity from hard
scatterings (eg. \cite{bas03}), because even hard partons can
participate significantly in equilibration processes, as is evidenced 
by the high-$p_{T}$ suppression found in Au + Au at RHIC.

Nonlinear effects are not considered here. Their
possible role in the context of relativistic
heavy-ion collisions has been discussed in \cite{lav02,ryb03,wol03}.
These account to some extent for the collective expansion of the
system in $y-$space, which is not included a priori in a statistical
treatment. In the linear model, the expansion is
treated through effective
diffusion coefficients $D_{y}^{eff}$ that are larger than the
theoretical values calculated from the dissipation-fluctuation theorem that 
normally relates $D_{y}$ and $\tau_{y}$ to each other \cite{wols99}.
One can then deduce the collective expansion velocities
from a comparison between data and theoretical result.  

The FPE can be solved analytically in the linear case  
with constant $D_{y}^{k}$. For net-baryon rapidity distributions,
the initial conditions are $\delta$-functions at the
beam rapidities $y_{b}=\pm y_{max}$. However, it has been shown that in
addition there exists a central (k=3, equilibrium) source at RHIC energies
which accounts for about 14{\%} of the net-proton yield in Au + Au
collisions at 200 AGeV \cite{wol03}, and is most likely related to
deconfinement. For d + Au, net-proton rapidity distributions are not
yet available.

For produced particles, the initial conditions are not uniquely
defined. Our previous experience with the Au + Au system
regarding both net baryons \cite{wol03}, and produced hadrons
\cite{biy04} favors a three-sources 
approach, with $\delta$-function initial conditions at the beam
rapidities, supplemented by a source centered at the equilibrium value
y$_{eq}$. This value is equal to zero
for symmetric systems, but for the asymmetric d + Au case its
deviation from zero according to (\ref{yeq}) is decisive 
in the description of particle production.

Physically, the particles in this source are expected to be generated
mostly from gluon-gluon collisions since only few valence quarks are
present in the midrapidity region at $\sqrt{s_{NN}}$ = 200 GeV
\cite{wol03}. Particle creation from a gluon-dominated source,
in addition to the sources related to the valence part of the 
nucleons, has also been proposed by Bialas and Czyz \cite{bia05}.
The final width of this source
corresponds to the local equilibrium temperature of the system which
may approximately be obtained from analyses of particle abundance ratios, plus
the broadening due to the collective expansion of the system.
Formally, the local equilibrium distribution is a solution
of (\ref{fpe}) with diffusion coefficient
$D_{y}^{3}$ = $D_{y}^{eq}$, and $\delta$-function initial condition at the
equilibrium value.
 
The PHOBOS-collaboration has analyzed their minimum-bias data 
successfully using a triple Gaussian fit
\cite{bbb04}. This is consistent with our analytical
three-sources approach, although additional
contributions to particle production have been proposed
\cite{liu04}. Beyond the precise representation of the data, however,
the Relativistic Diffusion Model offers an analytical description of
the statistical equilibration during the collision and in particular, 
of the extent of the moving midrapidity source which is indicative
of a locally equilibrated parton plasma prior to hadronization.

With $\delta-$function initial conditions for the Au-like source (1),
the d-like source (2) and the equilibrium source (eq), we obtain 
exact analytical diffusion-model solutions as an incoherent
superposition of the distribution functions $R_{k}(y,t)$ because the
differential equation is linear. The three individual distributions
are Gaussians with mean values
\begin{equation}
<y_{1,2}(t)>=y_{eq}[1-exp(-t/\tau_{y})] \mp y_{max}\exp{(-t/\tau_{y})}.
\label{mean}
\end{equation}
for the sources (1) and (2), and $y_{eq}$ for the
moving equilibrium
source. Hence, all three mean values attain y$_{eq}(b)$ as determined
from (\ref{yeq}) for t$\rightarrow \infty$, whereas for short times
the mean rapidities are smaller than, but close to the Au- and
d-like values in the sources 1 and 2. The variances are
\begin{equation}
\sigma_{1,2,eq}^{2}(t)=D_{y}^{1,2,eq}\tau_{y}[1-\exp(-2t/\tau_{y})].
\label{var}
\end{equation}

The charged-particle distribution in rapidity space is then obtained
as incoherent 
superposition of nonequilibrium and local equilibrium solutions of
 (\ref{fpe}) 
\begin{equation}
\frac{dN_{ch}(y,t=\tau_{int})}{dy}=N_{ch}^{1}R_{1}(y,\tau_{int})+
N_{ch}^{2}R_{2}(y,\tau_{int})
+N_{ch}^{eq}R_{eq}^{loc}(y,\tau_{int})
\label{normloc1}
\end{equation}
with the interaction time $\tau_{int}$ (total integration time of the
differential equation). In the present work, the integration is 
stopped at the value of $\tau_{int}/\tau_{y}$ that produces the
minimum $\chi^{2}$ with respect to the data and hence, the
explicit value of $\tau_{int}$ is not needed as an input. 
The result for central collisions is $\tau_{int}/\tau_{y} 
\simeq 0.4$.

The average numbers of charged particles in
the Au- and d-like regions $N_{ch}^{1,2}$ are proportional to the respective
numbers of participants $N_{1,2}$,
\begin{equation}
N_{ch}^{1,2}=N_{1,2}\frac{(N_{ch}^{tot}-N_{ch}^{eq})}{(N_{1}+N_{2})}
\label{nch}
\end{equation}
with the constraint $N_{ch}^{tot}$ = $N_{ch}^1$ + $N_{ch}^{2}$ +
$N_{ch}^{eq}$.
Here the total number of charged particles in each centrality bin
$N_{ch}^{tot}$ is determined from the data. The average number
of charged particles in the equilibrium source $N_{ch}^{eq}$ is a
free parameter that is optimized together with the variances
and $\tau_{int}/\tau_{y}$ in a $\chi^{2}$-fit of the data
using the CERN minuit-code \cite{jam81}. With known $\tau_{int}$, 
including its dependence on centrality, one could then 
determine $\tau_y$ and $D_y$, but this is beyond the scope of the present work.

The result of the RDM calculation is shown in Fig. 1
for central collisions (0-20\%) of d + Au. The charged-particle yield is
dominated by hadrons produced from the Au-like source, but there
is a sizeable equilibrium source that is more important
than the d-like contribution. This thermalized source is moving since
y$_{eq}$ has a negative value for d + Au, whereas it is zero
for symmetric systems.

The total yield is
compared to PHOBOS data \cite{bbb05} which refer to the
pseudorapidity $\eta=-ln[tan(\theta / 2)]$ since particle 
identification was not available.
As a consequence, there is a small difference to the model result
in $y$-space ($y\approx \eta$) which is most pronounced in the 
midrapidity region. It is removed when
the theoretical result is converted to $\eta$-space  
through the Jacobian
\begin{equation}
J(\eta,\langle m\rangle/\langle p_{T}\rangle) 
 = \cosh({\eta})\cdot [1+(\langle m\rangle/\langle p_{T}\rangle)^{2}
+\sinh^{2}(\eta)]^{-1/2}.
\label{jac}
\end{equation}
Here we approximate the average mass $<m>$ of produced charged hadrons in the
central region by the pion mass $m_{\pi}$, and use a
mean transverse momentum $<p_{T}>$ = 0.4 GeV/c. In the
Au-like region, the average mass is larger due to the
participant protons, but since their number $Z_{1}< 5.41$ is small compared to the 
number of produced charged hadrons in the d + Au system, the
increase above the pion mass remains small: 
$<m>\approx m_{p}\cdot Z_{1}/N_{ch}^{1} + m_{\pi}\cdot
(N_{ch}^1-Z_{1})/N_{ch}^1 \approx 0.17  GeV$. 
This increase turns out to have a negligeable effect on the results
of the numerical optimization, where we use $<m>/<p_{T}>=0.45$ for
the Jacobian transformations in the three regions. For reasonable
deviations of the mean transverse momentum from 0.4 GeV/c, the
results remain consistent with the data within the experimental error 
bars.

The equilibrium source in the light and asymmetric d + Au system
is found to contain only 19\% of the produced charged hadrons in 
central collisions. A previous result \cite{biy04} for Au + Au in the 
three-sources-RDM shows that the equilibrium source for 
particle production tends to be larger in the heavy
system \cite{biy04} at the same energy. Note, however, 
that the results of the $\chi^{2}-$minimization are not unique for Au 
+ Au due to the symmetry of the system.

The results for the mean values and variances of produced charged hadrons
in d + Au collisions
as functions of impact parameter are shown in Fig. 2. Here the
average impact parameters for the five centrality cuts $k$ are determined
according to $<b_{k}>=\int b\sigma_{k}(b)db/\int\sigma_{k}(b)db$
with the geometrical cross sections $\sigma_{k}(b)$ in each bin. Whereas
the total particle number and the particles created from
the Au-like source decrease almost linearly with increasing impact
parameter, the magnitude of the equilibrium source is roughly
independent of centrality. As a consequence, particle production
in the equilibrium source is relatively more important
in peripheral collisions. The variance of the central source
lies for sufficiently small impact parameters between the values for
the Au- and d-like sources.

The model calculations are converted to $\eta$-space and 
compared with PHOBOS data for 
five centrality cuts \cite{bbb05} and minimum bias \cite{bbb04}
in Fig. 3. The minimization procedure yields precise results
so that reliable values for the relative importance of the
three sources for particle production can be determined, Table 1.
The rapidity relaxation times and diffusion coefficients
can also be obtained from (\ref{mean}),(\ref{var}),
but this requires an independent information about the interaction
times. A small discrepancy in case of the most
peripheral collisions (80-100\%) is a consequence of 
the three straggling data points in the region -4 $<\eta<-3$.
The observed shift of the distributions towards
the Au-like region in more central collisions, and the steeper slope 
in the deuteron direction as compared to the gold direction
appear in the Relativistic Diffusion Model as a
consequence of the gradual approach to equilibrium.

Given the structure of the underlying differential equation
that we use to model the equilibration,
together with the initial conditions
and the constraints imposed by Eqs. ($\ref{yeq})$ and ($\ref{nch}$),
there is no room for substantial modifications of this result.
In particular, changes in the impact-parameter dependence of the mean 
values in (\ref{mean}) that are not in accordance with 
(\ref{yeq}) vitiate the precise agreement with the data.

To conclude, we have investigated 
charged-particle production in d + Au collisions at
$\sqrt{s_{NN}}$= 200 GeV as function of centrality
within the framework of an analytically soluble three-sources 
model. Excellent agreement with recent PHOBOS pseudorapidity
distributions has been obtained, and from a $\chi^{2}$-minimization we have 
determined the diffusion-model parameters
very accurately. 

For central d + Au collisions, a fraction of only 
19\% of the produced particles arises from the locally equilibrated
midrapidity source. Although this fraction increases towards more
peripheral collisions, the formation of a thermalized parton
plasma prior to hadronization can probably only be expected for
more central collisions.

The d + Au results show clearly that only the midrapidity part of the
distribution function reaches thermal equilibrium, whereas the
interaction time is too short for the d- and Au-like parts 
to attain the thermal limit. The same is true for the heavy Au + Au 
system at the same energy, but there the precise fraction of
particles produced in the equilibrium source is more
difficult to determine due to the symmetry of the problem.
 
One of the authors (GW) acknowledges the hospitality of the Faculty
of Sciences at Shinshu University, and financial support by
the Japan Society for the Promotion of Science (JSPS).

\newpage
\rm
Table 1. Produced charged hadrons as functions of
centrality in d + Au  
collisions at $\sqrt{s_{NN}}$ =
200 GeV, y$_{b}=\pm$ 5.36 in the Relativistic Diffusion Model.
The variance of the central source in $y-$space is $\sigma_{eq}^{2}$.
The number of produced charged particles 
is $N_{ch}^{1,2}$ for the sources 1 and 2 and $N_{ch}^{eq}$ for the equilibrium
source, the percentage of
charged particles produced in the thermalized source is $n_{ch}^{eq}$.
\\[1.5cm]
\begin{tabular}{lccccccc}
\hline
\hline
$Centrality  (\%)$ & 
$\sigma_{eq}^{2}$
&$N_{ch}^{1}$&$N_{ch}^{2}$&$N_{ch}^{eq}$&$n_{ch}^{eq}$(\%)&\\
\hline
0-20 &   3.99 & 131 &  19 &35&19\cr
20-40 &  3.95 & 78 &  17 &31&25\cr
40-60 &   5.70 &33 &  11 &38&46\cr
60-80 &   7.44 &9 &  5 &35&71\cr
80-100 &   6.89 & 2 &  2 &24&86\cr
min. bias& 4.04 & 56 & 15 & 21 & 23 \cr 
\hline
\hline
\end{tabular}
\newpage
\Large\bf
Figure captions
\normalsize\rm
\begin{description}
\item[Fig. 1.]
Charged-particle rapidity spectrum d$N_{ch}$/dy in the Relativistic Diffusion Model
(RDM) for central (0-20\%) d + Au collisions at $\sqrt{s_{NN}}$ = 200 GeV,
upper curve.  Dashed curves are contributions from the Au- and
d-like sources, respectively. The dashed area is the yield from the
moving thermalized central source of partonic origin, cf. text.
PHOBOS pseudorapidity data d$N_{ch}$/d$\eta$ ($\eta \approx y$) \cite{bbb05} are shown 
to illustrate the difference in $y-$ and $\eta-$space.
See Fig. 3 for a fit with the proper Jacobian transformation.
\item[Fig. 2.]
Mean values (upper part) and variances (lower part) of the three
sources for produced charged particles in rapidity space as functions of the
mean impact parameter for five centrality bins, symbols. Lines are drawn to guide 
the eye. Dashed lines have Au- and d-like sources, solid lines
correspond to the thermalized central source. The top solid
line gives the mean total number of produced charged particles.
\item[Fig. 3.]
Calculated pseudorapidity distributions of charged particles from
d + Au collisions at $\sqrt{s_{NN}}$ = 200 GeV for five different
collision centralities, and minimum-bias in comparison with
PHOBOS data \cite{bbb05,bbb04}. The analytical RDM-solutions are
optimized in a fit to the data. The corresponding 
minimum $\chi^{2}$-values (top left to bottom right) are 4.7, 5.9,
2.4, 1.7, 1.9, 2.1.
See Fig. 2 and text for the resulting diffusion-model parameters.
\end{description}
\newpage
\vspace{1cm}
\includegraphics[width=12cm]{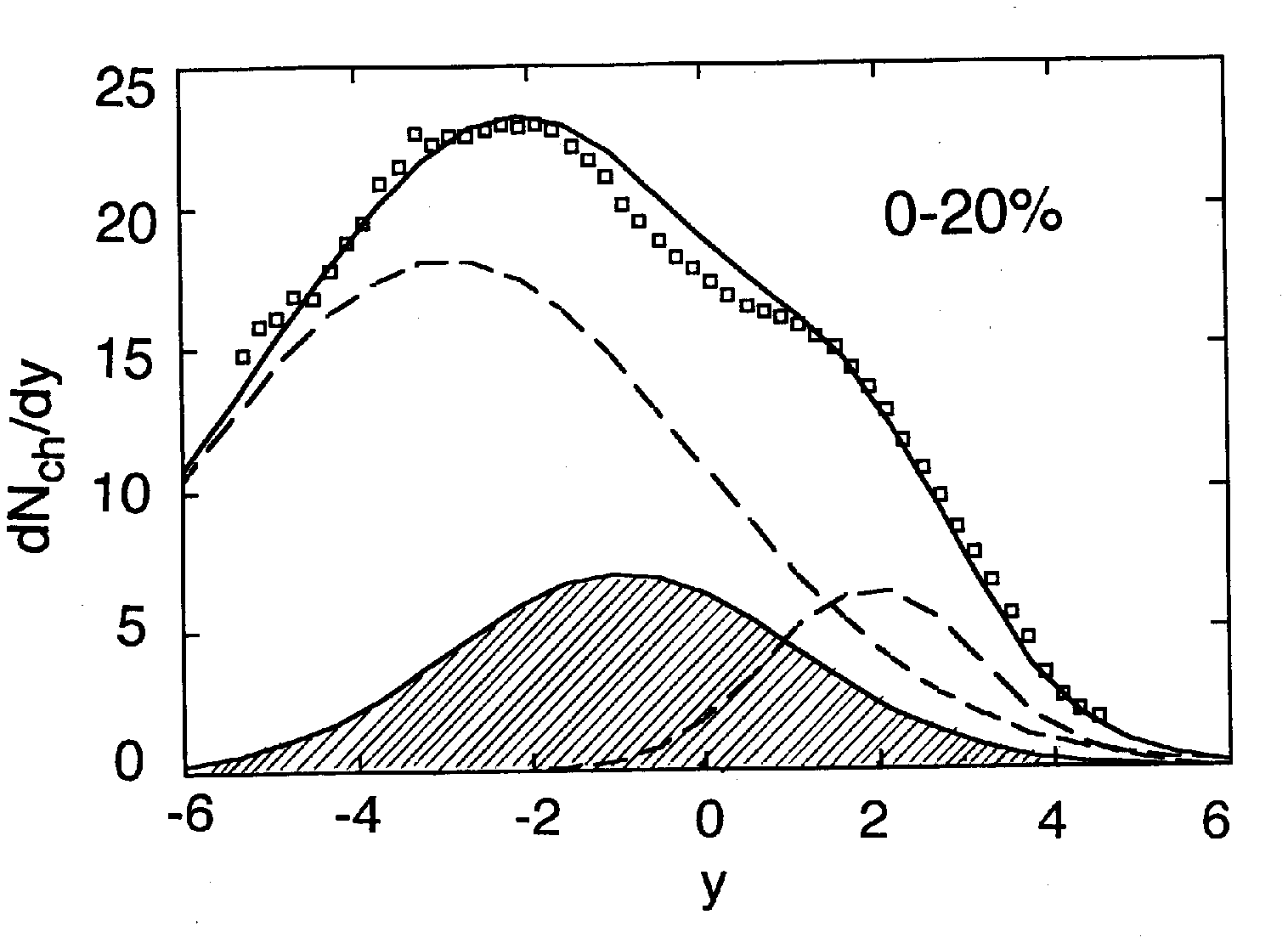}
\newpage
\vspace{1cm}
\includegraphics[width=12cm]{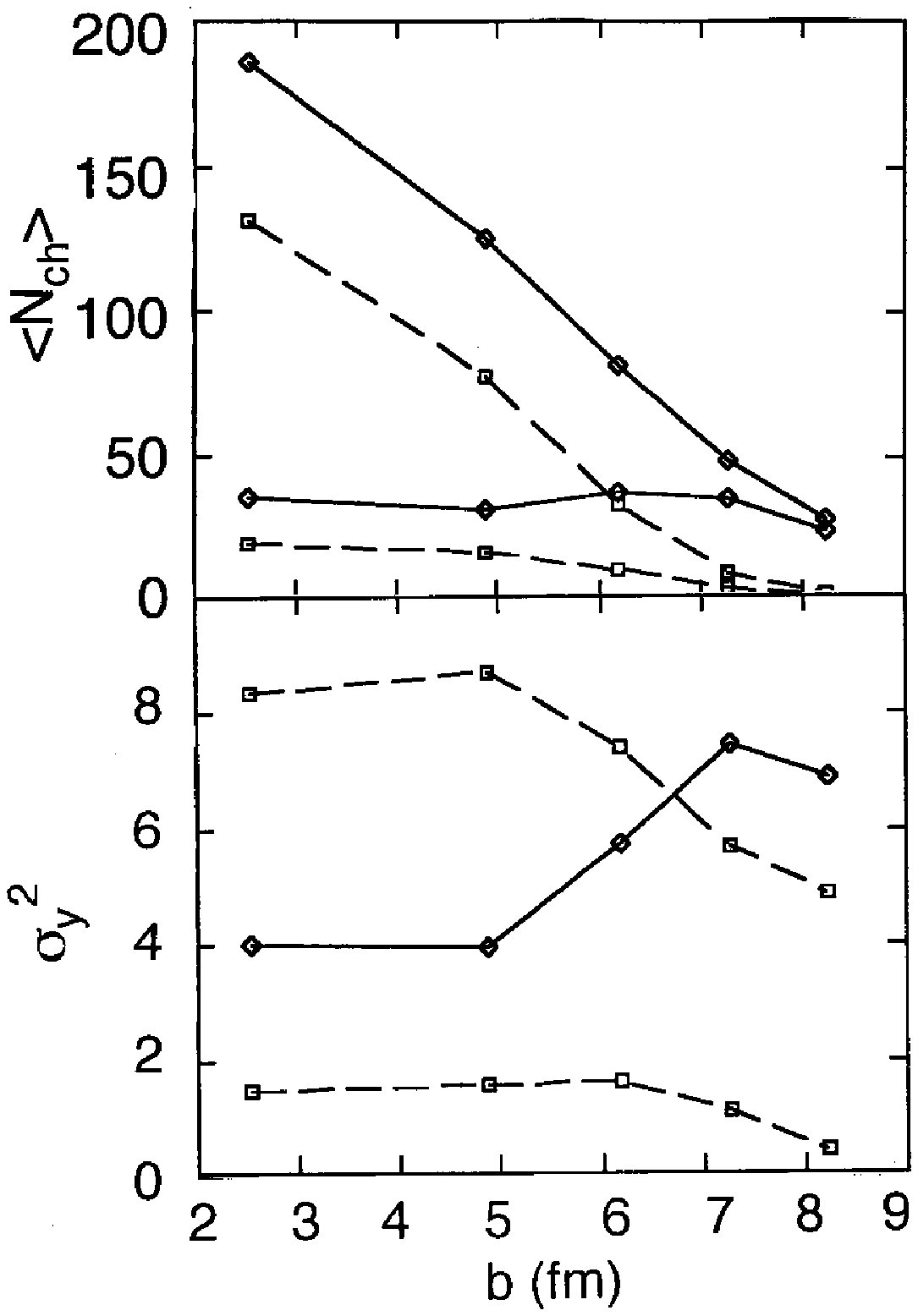}
\newpage
\vspace{1cm}
\includegraphics[width=15.2cm]{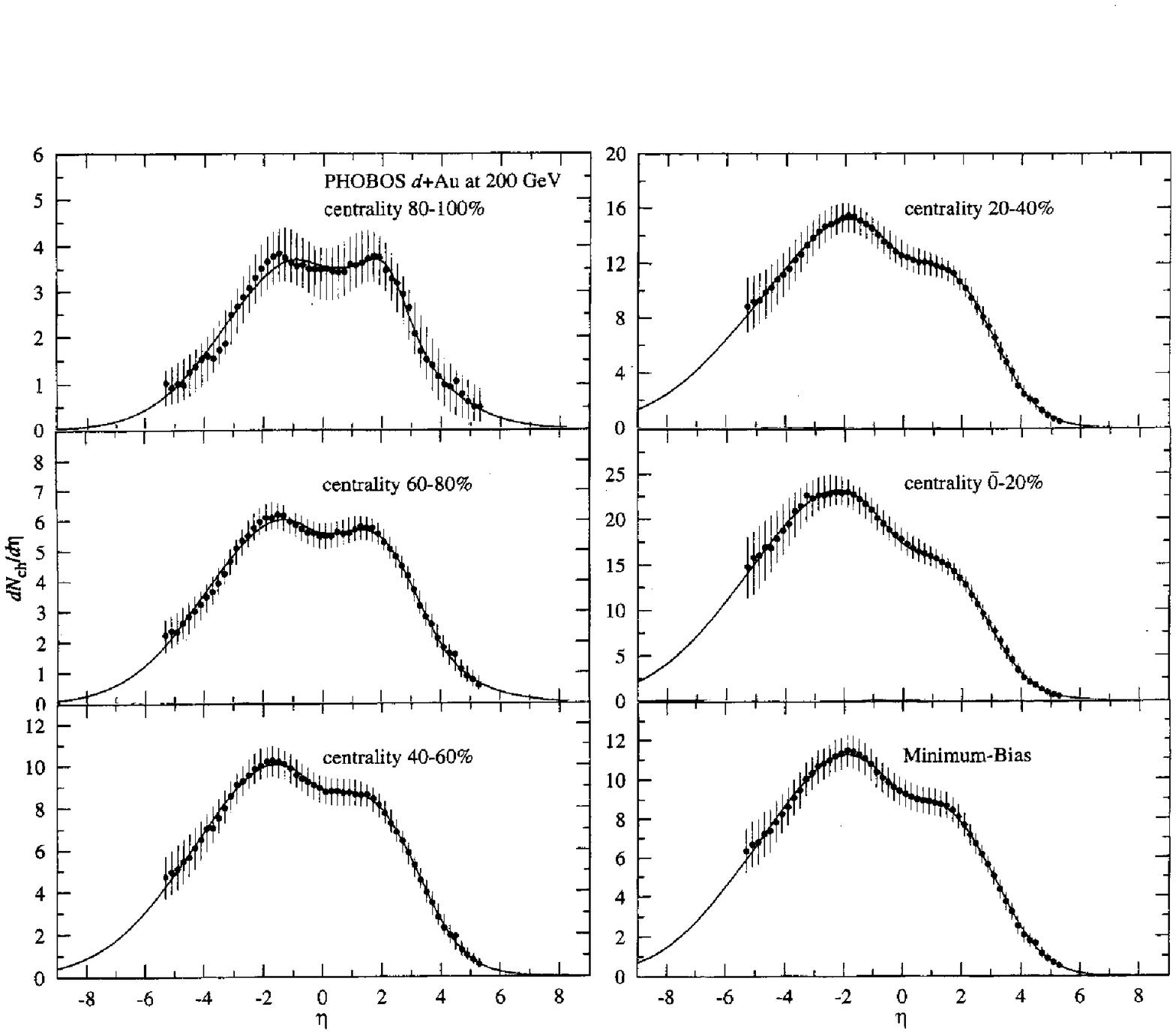}
\end{document}